\documentclass[%
 reprint,
 superscriptaddress,
 amsmath,amssymb,
 aps,
 pra,
 showkeys,
 floatfix,
]{revtex4-2}

\usepackage{multirow} 
\usepackage[utf8]{inputenc}
\usepackage[T1]{fontenc}
\usepackage{graphicx}
\usepackage{dcolumn}
\usepackage{bm}
\usepackage{booktabs}
\usepackage{array}
\usepackage[colorlinks=true, linkcolor=blue, citecolor=blue, urlcolor=blue]{hyperref}
\usepackage[capitalise]{cleveref}
\usepackage[all]{hypcap}
\usepackage{booktabs} 
\usepackage{amsmath}
\usepackage[table,xcdraw]{xcolor}
\usepackage{amssymb}
\usepackage{mathtools}
\usepackage{mathrsfs}
\usepackage{amsthm}
\usepackage{amssymb}
\usepackage{url}
\usepackage{makecell}
\usepackage{xcolor,soul}
\sethlcolor{yellow}
\usepackage[ruled,vlined,linesnumbered]{algorithm2e}
\soulregister\cite7
\usepackage{float}
\usepackage{siunitx}

\begin{document}

\preprint{APS/123-QED}

\title{Utility of NISQ devices: optimizing experimental parameters for the fabrication of Au atomic junction using gate-based quantum computers}
\author{Takumi Kanezashi}
\author{Daisuke Tsukayama}
\author{Jun-ichi Shirakashi}
\email{shrakash@cc.tuat.ac.jp}
\affiliation{Department of Electrical Engineering and Computer Science, Tokyo University of Agriculture and Technology, Koganei, Tokyo 184-8588, Japan}
\author{Tetsuo Shibuya}
\affiliation{Division of Medical Data Informatics, Human Genome Center, The Institute of Medical Science, The University of Tokyo, Minato, Tokyo 108-8639, Japan}
\author{Hiroshi Imai}
\affiliation{Graduate School of Information Science and Technology, The University of Tokyo, Bunkyo, Tokyo 113-8656, Japan}

\date{\today}

\begin{abstract}
Feedback-controlled electromigration (FCE) enables precise regulation of atomic migration by carefully optimizing multiple experimental parameters. However, manually fine-tuning these parameters poses significant challenges. This study investigated the feasibility of autonomously fabricating Au atomic junctions through gate-based quantum computing using a noisy intermediate-scale quantum (NISQ) device, which effectively approximates solutions to combinatorial optimization problems. We compared the computational accuracy of the NISQ device against a previously reported D-Wave quantum annealer. The results indicate that the NISQ device achieved lower residual energies and produced higher-quality approximate solutions for large-scale problems than the quantum annealing system.
\end{abstract}

\maketitle

Electromigration (EM) occurs when electrons flowing through a metallic nanowire collide with metal atoms, which induces atomic migration \cite{huntington1961current, black1969ieee, ho1989rep, etzion1975study}. Feedback-controlled electromigration (FCE) regulates EM intensity by applying voltage feedback (FB) control to Au nanowires, enabling the fabrication of Au atomic junctions using only electrical currents \cite{strachan2005controlled, campbell2013feedback, kanamaru2015ultrafast}. This method enables precise control of atomic migration at the individual level, facilitating the creation of nanoscale gaps for applications such as single-electron transistors \cite{bolotin2004metal, arzubiaga2014situ}, single-molecule transistors \cite{rattalino2012nanogap, perrin2015single}, and atomic junctions for nanoscale switches \cite{johnson2010memristive, wang2016single, schirm2013current}. We successfully fabricated single-electron transistors using FCE \cite{ito2018simultaneous}. Moreover, FCE can be extended to fabricate superconducting qubits when applied to superconducting metals \cite{nakamura1999coherent, pashkin2003quantum}. Achieving precise atomic control through FCE requires careful optimization of several experimental parameters, including feedback voltage $V_{\text{FB}}$, threshold differential conductance $G_{\text{TH}}$, and voltage step size $V_{\text{STEP}}$ \cite{kanamaru2015ultrafast, itami2010influence}. Scheduling optimal combinations of these parameters as experiments progress constitutes a combinatorial optimization problem. Previous studies leveraged autonomous systems incorporating machine learning \cite{iwata2020machine}, Ising computation \cite{sakai2019fabrication}, and quantum annealing (QA) \cite{yoneda2023searching} to optimize and schedule FCE parameters. Sakai et al. demonstrated improved atomic control in FCE experiments by optimizing these parameters using an Ising computing system \cite{sakai2019fabrication}. Similarly, Yoneda et al. employed QA to optimize experimental parameters within 20 ns, showcasing the rapid efficacy of QA \cite{yoneda2023searching}.

This study examined the potential of gate-based quantum computers in addressing this optimization problem. Contemporary gate-based quantum computers, known as noisy intermediate-scale quantum (NISQ) devices \cite{preskill2018quantum}, face challenges such as limitations in the number of physical qubits, coherence times, and available quantum gates. Nevertheless, advancements in quantum computing have increased qubit counts to over 100, and quantum superiority over classical methods has been experimentally validated \cite{bravyi2018quantum, bravyi2020quantum, kim2023evidence}.

This study investigated the feasibility of using gate-based quantum computers, specifically NISQ devices, for the autonomous optimization and scheduling of FCE experimental parameters, comparing their computational accuracy with that of previously reported QA systems.

Variational quantum algorithms (VQAs) \cite{cerezo2021variational} are widely recognized as effective methods for obtaining approximate solutions on NISQ devices. Notable examples include the variational quantum eigensolver (VQE) \cite{peruzzo2014variational, mcclean2016theory} and the quantum approximate optimization algorithm (QAOA) \cite{farhi2014quantum, farhi2017quantum}. These algorithms use the variational principle to construct quantum states with parameterized quantum circuits $U(\bm{\theta})$, forming an Ansatz:
\begin{equation}
|\psi(\bm{\theta})\rangle = U(\bm{\theta})|0\rangle.
\label{eq:Equation1}
\end{equation}

\begin{figure*}[t]
\centering
\includegraphics[width=\textwidth]{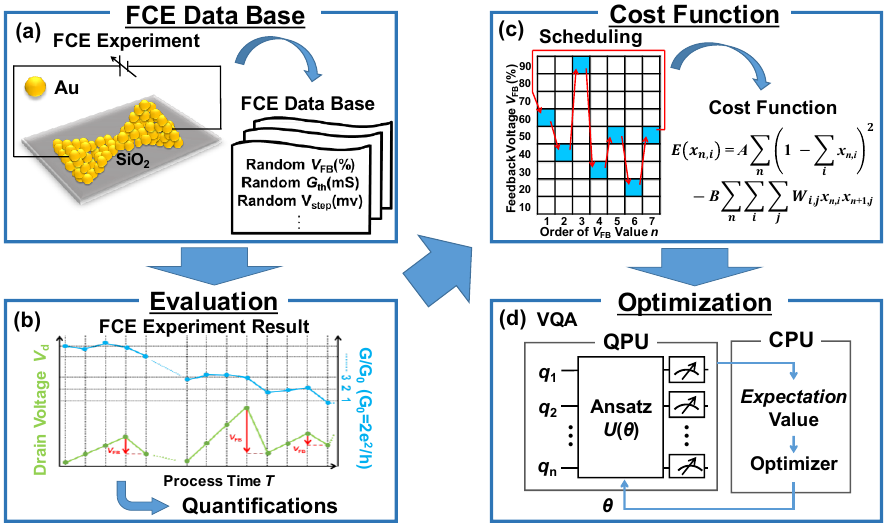}
\caption{Overview of the autonomous system for optimizing and scheduling FCE experimental parameters using a gate-based quantum computer. (a) A database is constructed based on experimental results obtained using randomly selected experimental parameters in FCE experiments. (b) The controllability of normalized conductance is evaluated under randomly applied experimental parameters. (c) The selection and scheduling of experimental parameters are formulated using one-hot encoding. (d) A ground-state search is conducted using a Variational Quantum Algorithm (VQA).}
\label{fig:Figure1}
\end{figure*}

The expectation value of the quantum state is calculated and passed to a classical optimizer, which updates the parameters $\bm{\theta}$ to minimize the expectation value. This iterative process continues until convergence, yielding the optimal parameters. The procedure accurately determines the ground-state energy and the corresponding wavefunction. The VQE and QAOA aim to solve the following optimization problem:
\begin{equation}
\min_{\bm{\theta}} \langle \psi(\bm{\theta}) | H | \psi(\bm{\theta}) \rangle,
\label{eq:Equation2}
\end{equation}
where $H$ denotes the Hamiltonian. Previous studies indicated that VQE produces more efficient and reliable approximate solutions than QAOA on NISQ devices \cite{miki2022variational}. Additionally, Nannicini reported that entanglement advantages are not observed in combinatorial optimization problems \cite{nannicini2019performance}. Based on these findings, we employed a hardware-efficient Ansatz using only $R_{\text{Y}}$ gate \cite{IBMQuantum} for each qubit. This approach reduced the number of optimization parameters and quantum gate operations, improving computational efficiency while mitigating gate errors and decoherence. The simplicity and low resource environments of $R_{\text{Y}}$ gates make them advantageous for NISQ devices with limited coherence times and gate fidelities. These advantages were evaluated by adopting a single $R_{\text{Y}}$-gate Ansatz and implementing an autonomous system driven by the VQE to optimize and schedule experimental parameters for FCE. This method enhances the precision and reliability of FCE parameter selection and scheduling while addressing the constraints of current NISQ hardware.

This study focused on optimizing $V_{\text{FB}}$, a critical experimental parameter in the FCE method. The voltage applied to Au nanowires incrementally increased during the FCE process, inducing EM that drives atomic migration over the feedback cycles. This process gradually reduces normalized conductance $G/G_0$, which approximates the number of atoms at the narrowest cross-section of Au atomic junctions \cite{umeno2009nonthermal, umeno2010spectroscopic}. The normalized conductance $G/G_0$ is defined as the conductance $G$ divided by the quantum conductance $G_0$ ($G_0 = 2e^2/h = \SI{77.6}{\micro\siemens}$, where $e$ denotes the electron charge and $h$ is Planck’s constant), and represents the fundamental unit of conductance in quantum systems. The FCE method prevents the excessive EM-induced breakage of Au nanowires by applying rapid feedback control to $V_{\text{FB}}$ when a steep decrease in normalized conductance is detected, effectively limiting atomic migration. Therefore, $V_{\text{FB}}$ plays a pivotal role in controlling quantum conductance.

Figure~\ref{fig:Figure1} illustrates the system flow of an autonomous method for scheduling optimal combinations of FCE experimental parameters using a gate-based quantum computer. Initially, FCE experiments were conducted using random $V_{\text{FB}}$ values, and a database was constructed based on the experimental outcomes (Fig.~\ref{fig:Figure1}(a)). The database, comprising 50 experimental datasets, aligns with prior studies. We evaluated the performance of the experimental parameters by analyzing the conductance curves within this database (Fig.~\ref{fig:Figure1}(b)). Figure 2(a) in Ref. \citenum{sakai2019fabrication} depicts a schematic of the normalized conductance $G/G_0$ and applied voltage $V$ as functions of process time $t$. To assess the normalized conductance, we introduce five variables: $D$, $F$, $L$, $P_1$, and $P_2$. Here, $D$ represents the decrease in normalized conductance during FB control, $F$ denotes the decrease in a single FB cycle, $P_1$ is the number of data points whose normalized conductance lies within $\pm 0.5 G_0$ of the reference value $G_{\text{ref}}$ immediately after FB, $P_2$ is the difference between the maximum $G_{\text{max}}$ and minimum $G_{\text{min}}$ values of the normalized conductance from immediately after FB until the end of the FB cycle, and $L$ is the total number of data points within this interval. Thus, $D$ and $F$ evaluate the degree of atomic displacement within the FB cycle, whereas $L$, $P_1$, and $P_2$ quantify the stability of the plateau in normalized conductance after FB. The score for the $n$-th $V_{\text{FB}}$ is defined as:
\begin{equation}
Score_{\text{VFB}} \equiv S_{\text{n}}(V_{\text{FB}}) = \frac{P_{1} / L}{|D - 1| + |F - 1| + |P_{2}|}.
\label{eq:Equation3}
\end{equation}
Higher $Score_{\text{VFB}}$ values correspond to smooth plateaus and stepwise decreases in quantum conductance $G_0$, which indicates stable one-by-one atomic migration in Au nanowires. Furthermore, to evaluate the relationship between the previous $V_{\text{FBi}}$ and subsequent $V_{\text{FBj}}$ voltages, we define $Score_{\text{trans}}$ as:
\begin{equation}
\begin{split}
Score_{\text{trans}} &\equiv S_{\text{n,n+1}}(V_{\text{FBi}}, V_{\text{FBj}}) \\
&= \frac{S_{\text{n}}(V_{\text{FBi}}) + S_{\text{n+1}}(V_{\text{FBj}})}{2},
\end{split}
\label{eq:Equation4}
\end{equation}
where $i, j \in \{1, \dots, Z\}$, with $Z$ denoting the number of $V_{\text{FB}}$ levels. The rounded interaction coefficients $W_{\text{ij}}$ were calculated by rounding the $Score_{\text{trans}}$ values (Fig. 2(c) in Ref. \citenum{sakai2019fabrication}) to a discrete set $\{0, 10, …, 90, 99\}$ using 115,239 data points across 511 feedback cycles. The $W_{\text{ij}}$ values from Ref. \citenum{yoneda2023searching} were incorporated to enable comparisons with QA results. A high $W_{\text{ij}}$ value indicates that the selected combination of experimental parameters before and after leads to a high degree of atomic mobility control. 

\begin{figure}[t]
\centering
\includegraphics[width=\columnwidth]{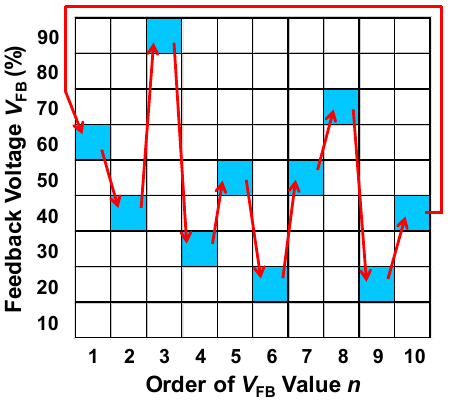}
\caption{Experimental parameter ($V_{\text{FB}}$) selection and scheduling using one-hot encoding, where one $V_{\text{FB}}$ is selected per order $n$.}
\label{fig:Figure2}
\end{figure}

A cost function was formulated using $W_{\text{ij}}$ to maximize the total score of the $V_{\text{FB}}$ schedule (Fig.~\ref{fig:Figure1}(c)). Figure 2 in Ref. \citenum{yoneda2023searching} represents the $V_{\text{FB}}$ schedule as variable $\bm{x}$ on an $N \times Z$ grid, where each element $x_{\text{n, i}}$ is a binary variable $\{0, 1\}$, with $N$ denoting the number of orders and $n$ representing the index of the $V_{\text{FB}}$ value in the schedule ($n \in \{1, \dots, N\}$). The mapped $V_{\text{FB}}$ schedule is expressed as a trajectory where $x_{\text{n, i}} = 1$. The number of nodes $x_{\text{n, i}}$ corresponds to the number of logical qubits employed. For example, Fig.~\ref{fig:Figure2} illustrates the use of 90 logical qubits arranged in a $9 \times 10$ configuration, cycling through the $V_{\text{FB}}$ schedule as follows: 60\% → 40\% → 90\% → 30\% → 50\% → 20\% → 50\% → 70\% → 20\% → 40\%. The cost function for optimizing the $V_{\text{FB}}$ schedule is defined as:
\begin{equation}
\begin{split}
E(\bm{x}) &= A \sum_{\text{n}} \left( \sum_{\text{i}} x_{\text{n,i}} - 1 \right)^{2} \\
&\quad - B \sum_{\text{n}} \sum_{\text{i}} \sum_{\text{j}} W_{\text{ij}} x_{\text{n,i}} x_{\text{n+1,j}} ,
\end{split}
\label{eq:Equation5}
\end{equation}
where $A$ and $B$ are hyperparameters. The first term enforces a one-hot encoding constraint, ensuring that only one $V_{\text{FB}}$ parameter is selected for each order $n$. The second term maximizes the total transition score of the $V_{\text{FB}}$ schedule. Consequently, maximizing the total score is equivalent to minimizing the cost function $E(\bm{x})$. This cost function follows the mathematical framework of the traveling salesman problem (TSP). Selecting experimental parameters based on the experiment’s progress is equivalent to sequentially selecting cities in the TSP. However, unlike the TSP, where “the same city cannot be visited twice,” Eq. (\ref{eq:Equation5}) does not impose a comparable constraint, namely, that “the same experimental parameter cannot be selected twice.” Consequently, this formulation has one fewer constraint than the standard TSP cost function. In the case of symmetric TSP, the distance matrix representing intercity distances is symmetric, with diagonal elements equal to zero. Our $W_{\text{ij}}$ for experimental parameter optimization is inspired by this distance matrix; however, as shown in Fig. 1(b) of Ref. \citenum{yoneda2023searching}, $W_{\text{ij}}$ is not symmetric. Furthermore, $W_{\text{ij}}$ varies depending on the experimental database used, indicating that our approach does not merely solve a benchmark problem but addresses complex real-world challenges. If the VQE on NISQ devices optimizes the FCE experimental parameters, it suggests the capability of the algorithm to solve complex combinatorial optimization problems, including the TSP. The ground-state search for the cost function $E(\bm{x})$ in Eq. (\ref{eq:Equation5}) was performed using the VQE (Fig.~\ref{fig:Figure1}(d)).

The computational performance of the VQE algorithm for optimizing the $V_{\text{FB}}$ schedule was evaluated using Eq. (\ref{eq:Equation5}). Solutions were obtained using the aer\_simulator\_matrix\_product\_state, a large-scale quantum simulator available in IBM's Qiskit, and two NISQ devices featuring the 127-qubit Eagle processor: ibm\_nazca and ibm\_brussels \cite{IBMQuantum}. The ibm\_nazca device used qubits with readout error below 5\% and no quantum circuit optimization. Conversely, ibm\_brussels applied quantum circuit optimization via Qiskit's generate\_preset\_pass\_manager \cite{IBMQuantum} with optimization level 3, prioritizing qubits exhibiting lower readout errors. Ground-state searches were conducted using 20 randomly initialized parameter sets for the simulator, four for ibm\_nazca, and five for ibm\_brussels. During the experiments, ibm\_nazca experienced over 5000 pending jobs, increasing wait times and lowering job priorities, which triggered adjustments to the experimental conditions. Each quantum circuit was executed with 8192 shots, with an average QPU usage time of approximately 6 s per circuit for both devices. Experiments used $Z = 9$ gradations for $V_{\text{FB}}$, with the number of orders $N$ ranging from 2 to 10, corresponding to logical qubit counts between 18 and 90. Experiments on ibm\_nazca were limited to $N \le 9$ due to time constraints. The classical optimizer NFT \cite{nakanishi2020sequential} was used to optimize the variational parameter $\bm{\theta}$, performing 1000 iterations to ensure convergence. Pre-testing the coefficients $A$ and $B$ in Eq. (\ref{eq:Equation5}) indicates that $A = 1000$ and $B = 7$ yielded the highest computational accuracy. These hyperparameters were adopted for subsequent gate-based quantum computing. QA experiments were conducted for comparison using D-Wave systems, including the D-Wave 2000Q, D-Wave Advantage, and D-Wave Advantage2 prototypes \cite{yoneda2023searching}. Due to the QA hardware structure, large-scale problems required logical qubits to be represented by groups of physical qubits or “chains,” a technique known as “minor embedding” \cite{choi2008minor, choi2011minor}. The minorminer module from the D-Wave Ocean SDK \cite{DWaveOceanSDK} facilitated this process. Experimental conditions for QA were consistent with those reported in Ref. \citenum{yoneda2023searching}.

\begin{figure}[t]
\centering
\includegraphics[width=\columnwidth]{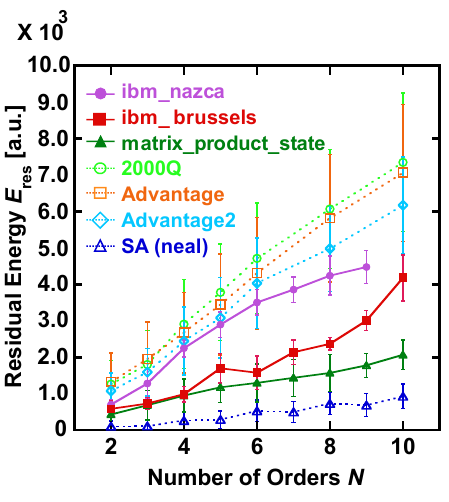}
\caption{Residual energy as a function of the number of orders $N$ for gate-based quantum computers, QA, and SA. Error bars represent standard deviations. Residual energy data for QA systems (D-Wave 2000Q, D-Wave Advantage, and D-Wave Advantage2) were obtained from Yoneda et al \cite{yoneda2023searching}.}
\label{fig:Figure3}
\end{figure}

\begin{table*}[t]
\centering
\caption{Logical qubit counts and average physical qubits used in QA systems (D-Wave 2000Q, Advantage, and Advantage2) as a function of the number of orders $N$. Physical qubit counts for gate-based quantum computers match logical qubit counts. QA systems require problem embedding into QPU qubits with limited connectivity, increasing physical qubit usage: Chimera topology (6-qubit connectivity) for D-Wave 2000Q, Pegasus topology (15-qubit connectivity) for D-Wave Advantage, and Zephyr topology (average 20 connections per qubit) for Advantage2.}
\label{tab:Table1}
\begin{tabular*}{\linewidth}{@{\extracolsep{\fill}} ccccc @{}}
\toprule
\makecell[c]{Number of\\orders} & \makecell[c]{Number of\\logical qubits} & \makecell[c]{Average number of\\physical qubits in\\D-Wave 2000Q} & \makecell[c]{Average number of\\physical qubits in\\D-Wave Advantage} & \makecell[c]{Average number of\\physical qubits in\\D-Wave Advantage2} \\
\midrule
2  & 18 & 105 & 47  & 41  \\
3  & 27 & 234 & 97  & 78  \\
4  & 36 & 340 & 134 & 112 \\
5  & 45 & 391 & 174 & 148 \\
6  & 54 & 450 & 208 & 174 \\
7  & 63 & N/A & N/A & N/A \\
8  & 72 & 687 & 286 & 233 \\
9  & 81 & N/A & N/A & N/A \\
10 & 90 & 849 & 374 & 295 \\
\bottomrule
\end{tabular*}
\end{table*}

Figure~\ref{fig:Figure3} illustrates the variation in residual energy across different backends as the number of orders $N$ increases. Determining ground states become increasingly challenging for larger combinatorial optimization problems. Consistent with a prior study \cite{yoneda2023searching}, the minimum energy $E_{\text{min\_SA}}$ obtained via simulated annealing (SA) using dwave-neal \cite{dwave_neal}, which demonstrated optimal performance on a CPU, was adopted as the reference solution. Residual energy $E_{\text{res}}$ is defined as:
\begin{equation}
E_{\text{res}} = \langle E \rangle - E_{\text{min\_SA}},
\label{eq:Equation6}
\end{equation}
As shown in Fig.~\ref{fig:Figure3}, $E_{\text{res}}$ increased with $N$ for gate-based quantum computers, QA, and SA, reflecting the growing difficulty of solving larger problems. Among these, $E_{\text{res}}$ was lowest for SA, followed by the gate-based quantum simulator, ibm\_brussels, ibm\_nazca, and QA. Notably, the NISQ devices ibm\_brussels and ibm\_nazca achieved lower residual energies than QA backends, which can be attributed to differences in the required number of physical qubits. Table~\ref{tab:Table1} lists the corresponding logical qubit counts, $N$, and the physical qubits used in QA. The number of physical qubits for the VQE equals the logical qubit count, while QA systems use minor embedding, where logical qubits are mapped onto physical qubits through a “chain,” significantly increasing qubit requirements: 5.8–9.4 times for D-Wave 2000Q, 2.6–4.2 times for D-Wave Advantage, and 2.3–3.3 times for D-Wave Advantage2. Yoneda et al. reported that QA, which uses fewer qubits, achieved higher solution accuracy even for the same problem size, with this tendency becoming more pronounced for larger problems. They attributed this to the occurrence of chain breaks as the chain lengthens, which leads to a decline in solution quality \cite{yoneda2023searching}. These findings suggest that NISQ devices require fewer qubits and experience less noise, even when employing the same superconducting architecture as QA systems, which enables them to more effectively suppress residual energy degradation. Among NISQ devices, ibm\_brussels achieved lower $E_{\text{res}}$ than ibm\_nazca, likely due to lower readout error rates. Remarkably, thanks to IBM’s advanced hardware capabilities and the quantum circuit optimization obtained from generate\_preset\_pass\_manager, the residual energy of ibm\_brussels was comparable to that of the noiseless simulator for $N \le 4$ using 36 qubits. The readout error rates for qubits used were 0.28–2.9\% for ibm\_brussels and 0.57–4.9\% for ibm\_nazca, with the 36th qubit at 1.2\%, the 56th at 1.7\%, and the 90th at 2.9\%. Efforts to mitigate readout errors, such as mthree \cite{nation2021scalable}, could bring $E_{\text{res}}$ closer to the simulator results, enabling more stable solutions.

\begin{figure*}[t]
\centering
\includegraphics[width=\textwidth]{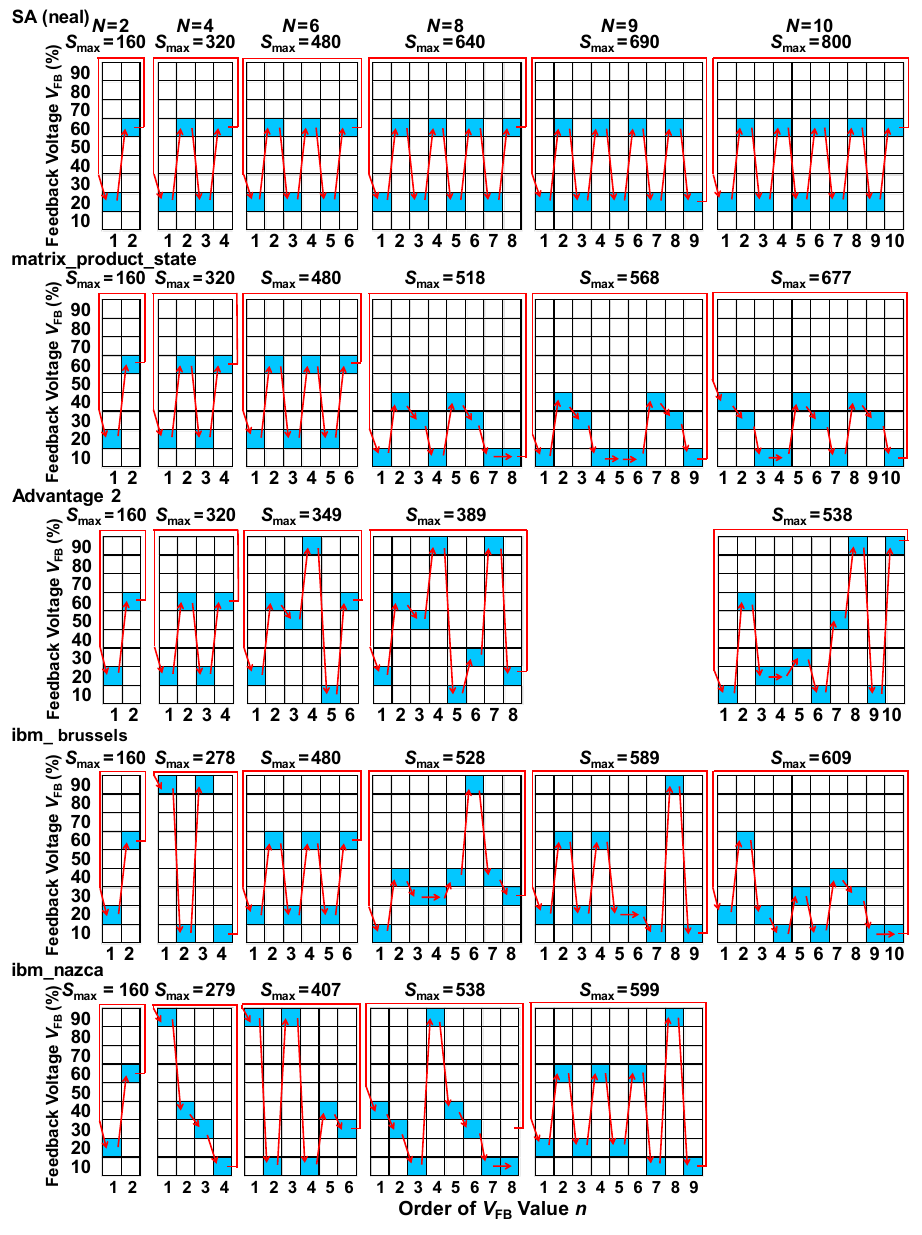}
\caption{$V_{\text{FB}}$ schedules obtained for each order $N$ using SA, aer\_simulator\_matrix\_product\_state, D-Wave Advantage2, ibm\_nazca, and ibm\_brussels, intended for practical FCE experiments.}
\label{fig:Figure4}
\end{figure*}

\begin{figure}[t]
\centering
\includegraphics[width=\columnwidth]{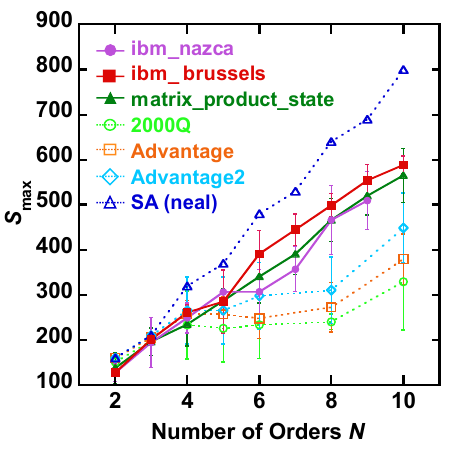}
\caption{Average $S_{\text{max}}$ values across trials as a function of the number of orders $N$ for gate-based quantum computers, QA, and SA. Trial counts: SA and QA (D-Wave 2000Q, D-Wave Advantage, and D-Wave Advantage2) performed 10 trials; aer\_simulator\_matrix\_product\_state, 20 trials; ibm\_nazca, 4 trials; ibm\_brussels, 5 trials. Error bars denote standard deviations.}
\label{fig:Figure5}
\end{figure}

The maximum $V_{\text{FB}}$ schedule was evaluated for solutions satisfying constraints using the metric $S_{\text{max}}$, defined as follows:
\begin{equation}
S_{\text{max}} = \sum_{\text{n}} \sum_{\text{i}} \sum_{\text{j}} W_{\text{ij}} x_{\text{n,i}} x_{\text{n+1,j}} ,
\label{eq:Equation7}
\end{equation}
which corresponds to the second term in Eq. (\ref{eq:Equation5}). Schedules with higher $S_{\text{max}}$ values were assumed to provide superior control over normalized conductance in FCE experiments. Figure~\ref{fig:Figure4} displays the highest $S_{\text{max}}$ schedules obtained using SA, aer\_simulator\_matrix\_product\_state, D-Wave Advantage2, ibm\_brussels, and ibm\_nazca. Schedules for $N =$ 2–6 from SA matched those from the simulator and were reproduced by D-Wave Advantage2 ($N = 2, 4$), ibm\_brussels ($N = 2, 6$), and ibm\_nazca ($N = 2$). Applying these schedules, alternating $V_{\text{FB}} =$ 20\% and 60\%, to FCE experiments confirmed excellent controllability of quantized conductance in Au atomic junctions \cite{sakai2019fabrication}. Partial repetition of the 20\% and 60\% $V_{\text{FB}}$ cycles was observed for $N =$ 9 and 10. Figure~\ref{fig:Figure5} depicts the variation in $S_{\text{max}}$ with increasing $N$ across different backends. Although no backend matched SA for $N \ge 4$ (logical qubits $\ge 36$), schedules from ibm\_brussels and ibm\_nazca achieved $S_{\text{max}}$ values comparable to the simulator. This suggests that, compared to the simulator, the quality of the schedules from NISQ devices remains high even when residual energy increases due to noise. The strong performance of these schedules indicates that they closely resemble the 20\% and 60\% repetitive schedules, which have been confirmed to exhibit high controllability over atomic movement.

Hence, current NISQ devices produce approximate solutions that are comparable to those of simulators. Gate-based quantum computers demonstrated high solution accuracy, matching that of QA for $N \le 4$ (logical qubits $\le 36$) and generating superior schedules compared to QA for $N \ge 5$ (logical qubits $\ge 45$). As previously noted, QA requires significantly more physical than logical qubits due to minor embedding, whereas gate-based devices maintain an equal number of logical and physical qubits. Consequently, gate-based quantum computers are less susceptible to noise, making their approximate solutions more accurate than those from quantum annealers.

This study evaluated the feasibility of autonomously optimizing FCE experimental parameters using gate-based quantum computers. Experimental parameter scheduling was framed as a combinatorial optimization problem. The schedules for small $N$ aligned with those previously identified through Ising spin computation, demonstrating effective control over normalized conductance \cite{sakai2019fabrication}. Partial matches were observed for $N = 9$ and $N = 10$, indicating high atomic controllability. The simulator, ibm\_nazca, and ibm\_brussels achieved lower residual energy than QA, with schedules comparable to those of the noiseless simulator. Notably, schedules outperformed QA-derived results for $N \ge 5$. These findings demonstrate that despite errors and limited coherence times, current NISQ devices can effectively optimize experimental parameters. Gate-based quantum computers exhibit significant potential for practical applications such as parameter optimization and scheduling, underscoring their relevance in addressing real-world challenges.

Future work will explore CVaR-VQE \cite{barkoutsos2020improving, chai2023optimal}, a method that converges faster and delivers improved solutions. The reduction in residual energy and the enhanced stability of solution discovery observed here, facilitated by selecting qubits with low readout errors, indicates potential gains through readout error mitigation techniques like mthree \cite{nation2021scalable}. Such methods could further reduce $E_{\text{res}}$, enabling more stable solutions. Additionally, higher-order binary optimization \cite{chai2023optimal, glos2022space}, which eliminates the constraint terms in Eq. (\ref{eq:Equation5}) to lower qubit usage, or the encoding scheme by Sciorilli et al. \cite{sciorilli2025towards}, which exponentially embeds logical qubits to solve larger problems with fewer physical qubits, could broaden the scope of solvable problems. These strategies enable the optimization of more experimental parameters, enhancing atomic migration control. This study demonstrates that the experimental parameters determined through an autonomous system using gate-based quantum computers effectively control atomic migration at the single-atom level. These findings mark a critical advancement toward leveraging gate-based quantum computers for autonomous systems and precise atomic-level control in practical applications.

\bibliography{reference}

\end{document}